\pdfoutput=1
\documentclass[a4paper,fleqn,usenatbib]{article}

\usepackage{graphicx}
\usepackage{amsmath}
\usepackage{subfig}
\usepackage{authblk}
\usepackage{natbib,hyperref}

\usepackage[a4paper,margin=2.5cm]{geometry}

\begin{document}
\title{Numerical methods for General Relativistic particles}
\author[1]{F. Bacchini\thanks{E-mail: fabio.bacchini@kuleuven.be}}
\author[1,2]{B. Ripperda}
\author[3]{L. Sironi}
\affil[1]{Centre for mathematical Plasma Astrophysics, Department of Mathematics, KU Leuven, Celestijnenlaan 200B, B-3001 Leuven, Belgium}
\affil[2]{Institut f\"{u}r Theoretische Physik, Max-von-Laue-Str. 1, D-60438 Frankfurt, Germany}
\affil[3]{Department of Astronomy, Columbia University, 550 W 120th St, New York, NY 10027, USA}
\renewcommand\Authands{ and }

\label{firstpage}
\maketitle

\begin{abstract}
We present recent developments on numerical algorithms for computing photon and particle trajectories in the surrounding of compact objects. Strong gravity around neutron stars or black holes causes relativistic effects on the motion of massive particles and distorts light rays due to gravitational lensing. Efficient numerical methods are required for solving the equations of motion and compute i) the black hole shadow obtained by tracing light rays from the object to a distant observer, and ii) obtain information on the dynamics of the plasma at the microscopic scale. Here, we present generalized algorithms capable of simulating ensembles of photons or massive particles in any spacetime, with the option of including external forces. The coupling of these tools with GRMHD simulations is the key point for obtaining insight on the complex dynamics of accretion disks and jets and for comparing simulations with upcoming observational results from the Event Horizon Telescope.
\end{abstract}

\section{Introduction and overview}
Particle-based numerical methods are a powerful tool in plasma physics. From collisionless magnetic reconnection (\citealt{sironispitkovsky2014}) to turbulent processes (\citealt{roytershteyn2015}), numerical methods such as Particle-in-Cell (PiC) have produced a variety of new physical results. In the astrophysical context, relativistic particle methods are used to model processes at the microscopic scale characterized by extremely high-energy, while the global scale is modeled with magnetohydrodynamics (MHD) approaches.

Current studies of such phenomena, when carried out with particles methods, are generally limited to locally flat regions of spacetime. Numerical schemes that take into account the underlying curvature of spacetime have not been actively developed (but see e.g. \citealt{levinsoncerutti2018}). Therefore, the microscopic dynamics of plasmas around compact objects such as black holes and neutron stars, where gravity plays an important role, is not well understood.

The physics of plasmas around black holes has received much attention in recent studies, thanks to ambitious projects that aim at the direct imaging of black holes for the first time (\citealt{falcke2017}). In this context, global simulations of black hole magnetospheres have been carried out with GRMHD approaches. Furthermore, from simulation data, synthetic radiation maps have been obtained, allowing for the reconstruction of the observed image as seen from a distant telescope. These images allow for direct comparison of observational data with theoretical predictions.

Obtaining such images involves a complex numerical approach that combines plasma physics on the global scale, optics, and microscopic processes. Here, advanced particle methods for general relativity play multiple roles: first, particle simulations can be used to obtain energy distributions from which the synthetic radiation maps are used. Current approaches generally assume Maxwellian or $\kappa$-distributions (\citealt{davelaar2018}). Second, once the radiation maps are available, rays of light must be traced from the physical point in space to the observer's position, in order to obtain an image that mirrors realistic observations. This step involves so called ``ray tracing'' techniques, which essentially consist of simulating bundles of photons (i.e. massless particles) traveling on geodesics.

Here, we aim both at improving existing particle methods used for simulating geodesic motion (for ray tracing), and at extending to the inclusion of the Lorentz force. This step is necessary in order to obtain particle methods that can ultimately be applied to the dynamics of plasmas around compact objects, in order to obtain realistic energy distributions and therefore more physically correct radiation maps used in the imaging process.

\section{General relativistic particles in curved spacetime}
The motion of particles subject to strong gravity is governed by the equation of motion
\begin{equation}
\frac{d^2x^{\mu}}{d\tau^2} + \Gamma^{\mu}_{\lambda\sigma} \frac{d x^{\lambda}}{d \tau} \frac{dx^{\sigma}}{d\tau} = qF^{\mu\nu}u_\nu,
\label{eq:geodesic}
\end{equation}
where $\mu=0,1,2,3$, for the four-position $x^\mu$ and an affine parameter $\tau$. The contravariant four-momentum is $g^{\mu\nu}u_\nu:=dx^\mu/d\tau$. $\Gamma^\mu_{\lambda\sigma}$ is the Christoffel symbol for the metric $g_{\mu\nu}$. In presence of electromagnetic fields, the force term on the right-hand side is given by the particle charge, $q$, and the Maxwell tensor, $F_{\mu\nu}=\partial_\mu A_\nu-\partial_\nu A_\mu$, where $A_\mu$ is the electromagnetic four-potential.

The equations above are typically written in the 3+1 language In this formulation, any metric of signature $(-,+,+,+)$ is given in the form
\begin{equation}
 g_{\mu\nu}=
 \begin{pmatrix}
  -\alpha^2+\beta_k\beta^k & \beta_i \\
  \beta_j & \gamma_{ij}
 \end{pmatrix},
 \label{eq:3p1metric}
\end{equation}
where $\alpha$ is the lapse function, $\beta^i$ is the shift three-vector, and $\gamma_{ij}$ is the spatial part of $g_{\mu\nu}$. Latin indices assume values from 1 to 3. With the definitions above, the equation of motion \eqref{eq:geodesic} can be rewritten in terms of first-order evolution equations in the variables $x^i$ and $u_i=g_{i\mu}u^\mu$, such that
\begin{equation}
\frac{d x^i}{dt} = \gamma^{ij} \frac{u_j}{u^0} - \beta^i,
\label{eq:geodesic3p1x}
\end{equation}
\begin{equation}
\frac{du_i}{dt} = -\alpha u^0 \partial_i \alpha + u_k \partial_i \beta^k - \frac{u_j u_k}{2 u^0} \partial_i \gamma^{jk} + q\left(F_{i0} + F_{ij}\frac{u^j}{u^0}\right),
\label{eq:geodesic3p1u}
\end{equation}
where 
\begin{equation}
u^0 = \left(\gamma^{jk} u_j u_k+\epsilon\right)^{1/2}/\alpha,
\label{eq:lfac}
\end{equation}
with $\epsilon=0$ for photons and $\epsilon=1$ for massive particles. In this formalism, the coordinate time $t:=x^0$ is used as affine parameter, and $u^0:=dt/d\tau$.

The system of equations (\ref{eq:geodesic3p1x}-\ref{eq:geodesic3p1u}) is suitable for numerical integration. It can be shown that the definition of $u^0$ above enforces the conservation of the norm of the four-velocity, $u^\mu u_\mu=-\epsilon$. Additionally, integrating in coordinate time $t$ rather than in proper time $\tau$ makes it easier to embed the motion of test particles in the time evolution of global electromagnetic fields.

In this work, we consider stationary metrics with no dependence on coordinate time $t$, hence the metric $g_{\mu\nu}$ depends on $x^i$ only. As a consequence, in all cases the energy, $E=-u_0$ is conserved. This physical property is often not respected during simulations, due to numerical errors affecting any discretization scheme. However, in some cases the lack of exact energy conservation can be detrimental for the accuracy of the results. Here, we present a new implicit numerical scheme that conserves energy to machine precision during the computation. The scheme is constructed by enforcing that the underlying Hamiltonian (i.e. the energy) is constant in time.

For simplicity, consider the case of pure geodesic motion, $A_\mu=0$. The Hamiltonian for stationary metrics is defined as
\begin{equation}
 H(x^a,u_b)=\alpha(\gamma^{jk} u_j u_k+\epsilon)^{1/2} - \beta^j u_j.
 \label{eq:hamiltonian}
\end{equation}
It is straightforward to derive the equations of motion by differentiating $H$, such that
\begin{equation}
 \frac{d x^i}{dt}=\frac{\partial H(x^a,u_b)}{\partial u_i}=\frac{\alpha \gamma^{ij} u_j}{(\gamma^{jk} u_j u_k+\epsilon)^{1/2}} -\beta^i,
\end{equation}
\begin{equation}
 \frac{d u_i}{dt}=-\frac{\partial H(x^a,u_b)}{\partial x^i}=-(\gamma^{jk} u_j u_k+\epsilon)^{1/2}\partial_i \alpha -\frac{1}{2}\frac{\alpha u_j u_k}{(\gamma^{jk} u_j u_k+\epsilon)^{1/2}}\partial_i \gamma^{jk} + u_j\partial_i \beta^j,
\end{equation}
which are precisely equations (\ref{eq:geodesic3p1x})-(\ref{eq:geodesic3p1u}) above. One can immediately verify that the corresponding discretized system
\begin{equation}
 \frac{\Delta x^i}{\Delta t}=\frac{\Delta^i_u H(x^a,u_b)}{\Delta u_i},
 \label{eq:deltax}
\end{equation}
\begin{equation}
 \frac{\Delta u_i}{\Delta t}=-\frac{\Delta_x^i H(x^a,u_b)}{\Delta x^i},
 \label{eq:deltau}
\end{equation}
satisfies the condition
\begin{equation}
 \frac{\Delta H(x^a,u_b)}{\Delta t} = \frac{\Delta^x_i H(x^a,u_b)}{\Delta  x^i}\frac{\Delta x^i}{\Delta t}+\frac{\Delta_u^i H(x^a,u_b)}{\Delta u_i}\frac{\Delta u_i}{\Delta t} = 0,
 \label{eq:hamenergyconddisc}
\end{equation}
and therefore it conserves energy. Here, $\Delta$ indicates a total discrete derivative, such that $\Delta H(x^a,u_b)=H(x^{a,n+1},u_b^{n+1})-H(x^{a,n},u_b^n)$. The operators $\Delta_x^i$ and $\Delta_i^u$, instead, correspond to discrete partial derivatives, namely with respect to $x^i$ and $u_i$. The exact form of the discrete operators can be found in full in \cite{bacchini2018a}.

A remarkable feature of this scheme is the absolute freedom in the definition of the Hamiltonian $H$. This implies that for systems characterized by a Hamiltonian different from that of equation ($\ref{eq:hamiltonian}$), the algorithm retains its energy (or in general, first integrals) conservation properties. Thus, the extension to more complicated physical situations becomes straightforward, provided that the corresponding Hamiltonian formulation is available. For this reason, it is immediate to include the Lorentz force, by employing the corresponding Hamiltonian,
\begin{equation}
 H(x^a,\pi_b) = \alpha\sqrt{1 + \gamma^{ij} \left(\pi_i-\frac{q}{m}A_i\right) \left(\pi_j-\frac{q}{m}A_j\right)} - \beta^k\left(\pi_k-\frac{q}{m}A_k\right) - qA_0,
 \label{eq:hamlorentz}
\end{equation}
where the conjugate three-momentum $\pi_i = u_i+qA_i$ appears in place of $u_i$ as an independent variable. The solution procedure remains unchanged with respect to the case of pure geodesic motion, hence making it possible to simulate charged particles in general relativistic contexts while retaining exact energy conservation.

\section{Results and outlook}
The numerical solution of equations \eqref{eq:deltax}-\eqref{eq:deltau} yields geodesic trajectories characterised by conservation of energy to machine precision. The advantage implied by this feature is two-fold. First, the motion of particles around metric singularities is particularly affected by numerical energy dissipation that characterizes the numerical schemes typically employed (e.g. explicit Runge-Kutta or implicit midpoint rule). The new ``Hamiltonian'' scheme, eliminates such errors and therefore does not require the extreme reduction of the time step demanded by other schemes. Second, since there is no secular growth of error in the energy, bound orbits confined within certain regions of space around a compact object can be simulated indefinitely without the spurious escape of the particle observed when using explicit methods.

\begin{figure}[!h]
\centering
\subfloat[Unstable spherical photon orbits around an extremal Kerr black hole.]{\includegraphics[scale=.238]{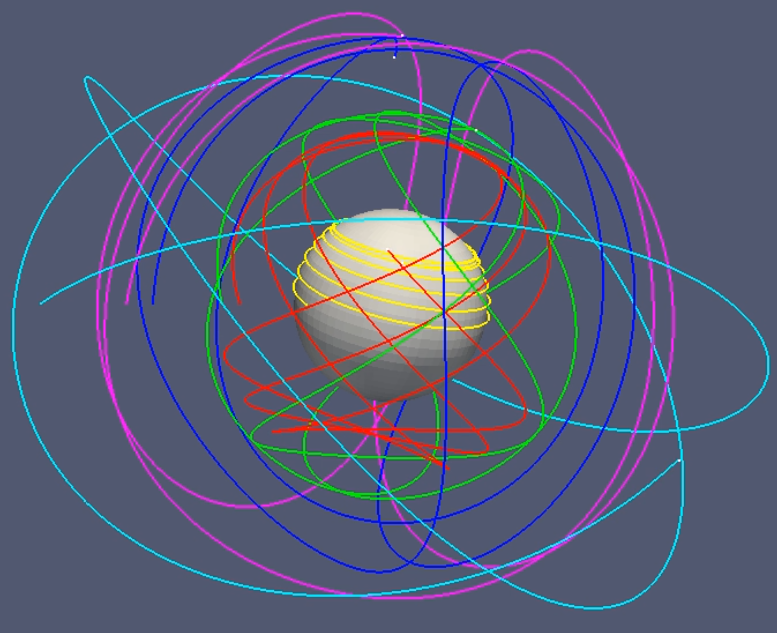}
\label{fig:phorbits}} \quad
\subfloat[Simulation of the observed shadow of an extremal Kerr black hole distorting the view of four-color background.]{\includegraphics[scale=.2]{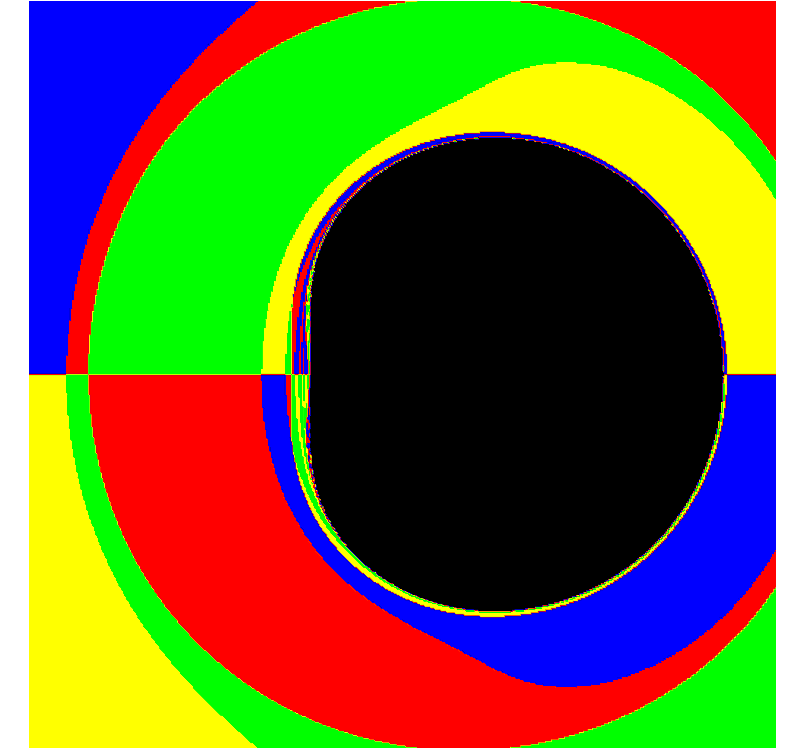}
\label{fig:bhshadow}}
\caption{Simulations of photon trajectories with the new Hamiltonian method.}
\end{figure}

The features above are exemplified in Figure \ref{fig:phorbits}, where several spherical photon orbits are shown around an extremal Kerr black hole. Such orbits are highly unstable, and cannot be easily simulated with standard methods unless the time step is greatly reduced. The Hamiltonian scheme, by contrast, inherently stabilizes the motion on the correct path. As a side effect, all other constants of the motion are exactly conserved (energy, angular momentum, and Carter constant).

For ray tracing simulations, the features above are especially attractive because they allow for the fast computation of so called ``black hole shadows'' via integration of a large number of geodesic paths. One example of such a calculation is shown in Figure \ref{fig:bhshadow}. Here, an extremal Kerr black hole is placed between a distant observer and a four-color background. The distortion of spacetime caused by the presence of the black hole bends the light rays close to the event horizon, creating the asymmetrical shadow. Such a picture can be directly compared to upcoming observations of black holes, with the aim of confirming our current theoretical models. Hence, it is crucial that the numerical solution is as accurate as possible, which is ensured by numerical methods such as the Hamiltonian scheme here presented.

Finally, as mentioned above, the scheme is directly applicable to the motion of charged particles. This is a very fundamental step for obtaining accurate simulations of plasmas around compact object from the microscopic perspective. The results of such particle simulations can be used as a basis for the calculation of energy distributions from which synthetic radiation maps are drawn. This lays the ground for improving our understanding of the dynamics of energetic outflows and flares that are currently being measured (\citealt{genzel2003}). Additionally, when the full particle feedback to the electromagnetic fields (in the PiC fashion) is taken into account, previously unreachable energy scales can actually be explored. Since the development of e.g. kinetic waves and instabilities is expected to be quantitatively different under these conditions, numerical schemes such as those here presented are the ideal tool to enrich our knowledge of these complicated physical processes.

\bibliographystyle{apalike}

\begin{thebibliography}{}

\bibitem[Bacchini et~al., 2018]{bacchini2018a}
Bacchini, F., Ripperda, B., Chen, A., and Sironi, L. (2018).
\newblock Generalized, energy-conserving numerical simulations of particles in
  general relativity. {I}. time-like and null geodesics.
\newblock {\em ApJS}, 237, 6.

\bibitem[Davelaar et~al., 2018]{davelaar2018}
Davelaar, J., Mo\'{s}cibrodzka, M., Bronzwaer, T., and Falcke, H. (2018).
\newblock General relativistic magnetohydrodynamical $\kappa$-jet models for
  {Sagittarius A*}.
\newblock {\em MNRAS}, 612, A34.

\bibitem[Falcke, 2017]{falcke2017}
Falcke, H. (2017).
\newblock Imaging black holes: past, present and future.
\newblock {\em J. Phys.: Conf. Ser.}, 942, 012001.

\bibitem[{Genzel} et~al., 2003]{genzel2003}
{Genzel}, R., {Sch{\"o}del}, R., {Ott}, T., {Eckart}, A., {Alexander}, T.,
  {Lacombe}, F., {Rouan}, D., and {Aschenbach}, B. (2003).
\newblock {Near-infrared flares from accreting gas around the supermassive
  black hole at the Galactic Centre}.
\newblock {\em Nature}, 425, 934.

\bibitem[Levinson and Cerutti, 2018]{levinsoncerutti2018}
Levinson, A. and Cerutti, B. (2018).
\newblock Particle-in-cell simulations of pair discharges in a starved
  magnetosphere of a kerr black hole.
\newblock {\em A\&A}, 616, A184.

\bibitem[{Roytershteyn} et~al., 2015]{roytershteyn2015}
{Roytershteyn}, V., {Karimabadi}, H., {Omelchenko}, Y., and {Germaschewski}, K.
  (2015).
\newblock {Turbulence dissipation challenge: particle-in-cell simulations}.
\newblock {\em AGU Fall Meeting Abstracts}, pages SH11E--2417.

\bibitem[Sironi and Spitkovsky, 2014]{sironispitkovsky2014}
Sironi, L. and Spitkovsky, A. (2014).
\newblock Relativistic reconnection: an efficient source of non-thermal
  particles.
\newblock {\em ApJL}, 783, L21.

\end{thebibliography}

\label{lastpage}
\end{document}